# High Responsivity in Molecular Beam Epitaxy (MBE) grown β-Ga$_2$O$_3$ Metal Semiconductor Metal (MSM) Solar Blind Deep-UV Photodetector

Anamika Singh Pratiyush[1,a)], Sriram Krishnamoorthy[2,a)], Swanand Vishnu Solanke[1], Zhanbo Xia[2], Rangarajan Muralidharan[1], Siddharth Rajan[2], Digbijoy N. Nath[1,b)]

[1]Centre for Nano Science and Engineering (CeNSE),
Indian Institute of Science (IISc), Bangalore 560012
[2]Department of Electrical and Computer Engineering, The Ohio State University,
Columbus, OH, 43210

**Abstract:**

**In this report, we demonstrate high spectral responsivity (SR) in MBE grown epitaxial β-Ga$_2$O$_3$-based solar blind MSM photodetectors (PD). (-2 0 1)-oriented β-Ga$_2$O$_3$ thin film was grown by plasma-assisted MBE on c-plane sapphire substrates. MSM devices fabricated with Ni/Au contacts in an interdigitated geometry were found to exhibit peak SR > 1.5 A/W at 236-240 nm at a bias of 4 V with a UV to visible rejection ratio > 10$^5$. The devices exhibited very low dark current < 10 nA at 20 V and showed no persistent photoconductivity (PPC) as evident from the sharp transients with a photo-to-dark current ratio > 10$^3$. These results represent the state-of-art performance for MBE-grown β-Ga$_2$O$_3$ MSM solar blind detector.**

[a)] Anamika Singh Pratiyush and Sriram Krishnamoorthy have equally contributed for this work.
[b)] Corresponding author email: digbijoy@cense.iisc.ernet.in

β-Ga$_2$O$_3$ is an emerging wide band gap material (E$_G$ ~4.6 eV) which is at the focus of a rapidly-expanding device and materials community for its promise towards enabling next-generation high-power transistors[1–3] and deep UV solar blind detectors[4–11] towards strategic applications such as missile plume detection and bio-medical sensors. III-nitride alloys (AlGaN), which have been widely explored [12–16] for solar blind UV detection, suffer from lack of native substrates which is a major bottleneck to achieving superior material quality. In contrast, large area single crystal Ga$_2$O$_3$ substrates can be grown from the melt by various conventional crystal growth techniques [17,18] with much superior crystal quality providing an economic and performance advantage. Besides, the growth of high-quality Al$_x$Ga$_{1-x}$N with increasing Al mole-fraction required for deep UV detection poses severe challenges in epitaxy[19]. Epitaxially grown β-(Al,Ga)$_2$O$_3$ provides a promising platform to tailor hetero-structure engineering for high-performance and novel devices. In particular, MBE offers an approach to achieving growth of high quality films with low background impurity concentration, which is critical for low dark current in photodetectors. In this work, we report on the hetero-epitaxial MBE growth of (-201) β-Ga$_2$O$_3$ on sapphire substrates and the demonstration of metal-semiconductor-metal (MSM) photodetectors with high responsivity in conjunction with a very low dark current. These results demonstrate the promise of β-(Al,Ga)$_2$O$_3$ devices fabricated not only on bulk substrates, but also on relatively low-cost sapphire substrates.

β-Ga$_2$O$_3$ samples were grown on c-plane sapphire substrates using MBE equipped with Veeco Uni-bulb O$_2$ plasma source and standard effusion cell for gallium. Sapphire substrates were solvent cleaned, indium bonded to a silicon wafer and degassed in the buffer chamber at 400°C for 1 hour before loading the sample into the main growth chamber. Ga$_2$O$_3$ was grown for 3 hours at a substrate temperature of 700°C with a RF plasma power of 300W and a Ga flux of 1.5x10$^{-8}$ Torr measured using a beam flux monitor. X-ray diffraction data shown in Figure 1(a) indicated single phase (-201)-oriented β-Ga$_2$O$_3$ film, which is in agreement with previous reports [20–22]. The thickness of the β-Ga$_2$O$_3$ film was estimated to be 150 nm from X-ray reflectivity (XRR) measurement (Fig. 1(b)).

MSM photodetectors were fabricated on MBE-grown (-201) β-Ga$_2$O$_3$ film. After following the standard lithography procedures, Ni/Au (20 nm/70 nm) metal stack was deposited using e-beam evaporator for Schottky contacts. The metal electrodes had an interdigitated geometry with 30 fingers: 300 μm long, 4 μm wide, 8 μm finger spacing with an active area of 360 μm x 300 μm (as shown in inset to Fig. 1(b)).

The spectral responsivity (SR) of the MSM photodetectors was measured using Sciencetech® Inc. quantum efficiency (QE) setup fitted with a Xenon lamp (150 Watts), optical chopper and monochromator, while the current-voltage (I-V) and time-dependent photo-response were measured using Keithley® 2450 source-meter connected externally to the QE setup.

Fig. 2 shows the spectral responsivity versus wavelength (linear and log scale) at bias voltages of 4 V, 8 V, 12 V, and 16 V. The peak wavelength was found to be ~ 236 nm (5.62 eV) while the band edge was estimated from the peak value of the derivative of the curve (not shown) to be at 254 nm (4.88 eV). Peak responsivity values of 1.8 A/W and 3.3 A/W were measured at bias of 4 V and 16 V respectively, the highest among all MBE-grown Ga$_2$O$_3$-based MSM devices as well as one of the best among all detector architectures for epitaxial Ga$_2$O$_3$. A few groups have reported higher SR values than what is reported here[23,24]; however, in those cases, the responsivity values have not been experimentally measured, but are estimated or calculated from photo-current measurements. Visible rejection ratio was estimated by dividing the responsivity at 236 nm (peak response) to that at 450 nm. Excellent visible rejection ratio $> 10^5$ was observed, which is among the highest reported for epitaxial MBE grown $Ga_2O_3$-based detectors [7,23] and testifies the true solar-blind nature of the devices.

Fig. 3 (a) shows the photo and dark current-voltage (I-V) characteristics of the device (linear and log scale) at room temperature. The photocurrent was measured at the illumination of 236 nm. High photo-current (~ 4.9 µA at 20 V) and very low dark current (~ 4 nA at 20 V) were measured and photo-to-dark current ratio $> 10^3$ were obtained.

To evaluate the performance of the photodetector, time-dependent photo-response measurements were done at different bias voltages. Fig. 3 (b) illustrates the time-dependent photo-response of the β-$Ga_2O_3$ photodetector under the illumination of 236 nm at a bias voltage 20 V in log scale. The inset of the Fig. 3 (b) shows the rescaled transient plot in the linear scale. Under deep UV illumination, the photocurrent rises to a value of 14 µA; when the UV illumination is turned off, the current abruptly decreases to a value of ~7 nA. The on/off ratio of >1000, the highest for MBE grown β-$Ga_2O_3$-based detectors, was observed with no persistent photo conductivity (PPC). It was observed that the photo current at 20 V was higher during transient measurement compared to that in steady-state I-V. This can be attributed to the temporal dependence of photo-current due to possible trapping-de-trapping transients. This effect was found to be absent after the devices were passivated with 20 nm of Atomic Layer Deposited (ALD) $Al_2O_3$ at 250°C using Trimethyl-aluminium (TMAl) and $H_2O$ as precursors, indicating that surface related traps are responsible for this effect.

Fig. 4(a) illustrates the photo and dark I-V characteristics and Fig. 4(b) shows time-dependent photo-response at a bias of 20 V after the passivation of the devices respectively. The photo currents in steady-state as well as transient measurements were found to be similar ~ 4.6 µA (at 20 V) while the dark current was observed to be in the ~ nA range. Both the photo and the dark current decreased slightly after the passivation of the devices.

With an incident power of ~ 0.7 mW/cm$^2$ over a device active area of ~ 300 μm x 360 μm, the maximum photo current is estimated to be ~ 0.15 μA (Assuming the external quantum efficiency to be 100 %) whereas the measured photo current using sourcemeter is ~ 4.6 μA ( at 20 V). This is a clear indication of the gain in the devices. Hole trapping has been suggested as one of the several possible[25] gain mechanisms in MSM photodetectors. We predict that the high gain resulting in high responsivity in the devices is due to the localization of the self-trapped holes at the interface of the metal and the semiconductor [25]. The photo-generated electrons are swept away quickly by the field whereas the hole remains self-trapped in the depletion region, lowering the barrier height at the junction. To maintain charge neutrality more electrons flow from the metal side, increasing the measured photocurrent and hence the gain. The measured photo current ($I_{ph}^m$) consists of two components: an intrinsic bias-independent photo current $I_{ph}^0$, and the component arising due to the barrier lowering ($\Delta\Phi_B$) which enhances the current ($I_{ph}^m$), leading to gain. Thus we have [26]

$$I_{ph}^m = [e^{\frac{\Delta\phi_B}{KT}} - 1]I_{Dark} + I_{ph}^o \qquad . \qquad (1)$$

Using equation (1), the barrier lowering ($\Delta\Phi_B$) due to trapping of holes is estimated to be ~ 0.27 eV as shown in the inset of the Fig (4(a)). This is in good agreement with the value of $\Delta\Phi_B$ ~ 0.3 eV reported by Armstrong et al [25].

After the passivation of the devices, the rise time (10 % - 90 %) was found to be 3.33 seconds while the fall time (90 % - 10 %) was measured to be 0.4 second (Fig. 4(b)). Slightly higher rise time compared to fall time was always observed which could be possibly attributed to the slow release of carriers from deep states and/or due to the slow hole transport on account of very low hole mobility[27,28]. The abrupt fall in the transient can be attributed to the fast recombination by both band-to-band recombination and via recombination centers[29].

Fig. 5 illustrates the plot of the responsivity versus dark current for the state-of-the-art solar blind (230-290 nm) UV detectors reported [4-6,8,10-11,16,30–37] for the devices based on $Al_xGa_{1-x}N$ as well as on $\beta$-$Ga_2O_3$. The detectors reported in this work have excellent responsivity while maintaining a very low dark current for 230-240 nm range.

In conclusion, we demonstrate a high responsivity > 1.5 A/W at 4 V for deep UV detectors based on MBE grown $\beta$-$Ga_2O_3$ with low dark current < 10 nA (at 20 V) and high visible rejection ratio > $10^5$. We also demonstrate abrupt transients in photo current without any PPC effect and report the highest on/off photo-to-dark current ratio > 1000 among MBE grown $\beta$-$Ga_2O_3$ solar blind detectors of any architecture. The high gain mechanism is explained by the Schottky barrier lowering due to self-trapped hole in the depletion region.

This work is funded by Joint Advanced Technology Program (JATP), grant number JATPO152, and by the Department of Science and Technology (DST) under its Water Technology Initiative (WTI), grant number DSTO1519. We would also like to thank Micro and Nano Characterization Facility (MNCF) and NNFC staff at CeNSE, IISc for their help and support in carrying out his work. Z.X. and S.R. acknowledge funding from the U.S. Office of Naval Research EXEDE MURI (Program Manager: Dr. Brian Bennett). This work was supported in part by The Ohio State University Materials Research Seed Grant Program, funded by the Center for Emergent Materials, an NSF-MRSEC, grant DMR-1420451, the Center for Exploration of Novel Complex Materials, and the Institute for Materials Research.

**Figure and legends**:

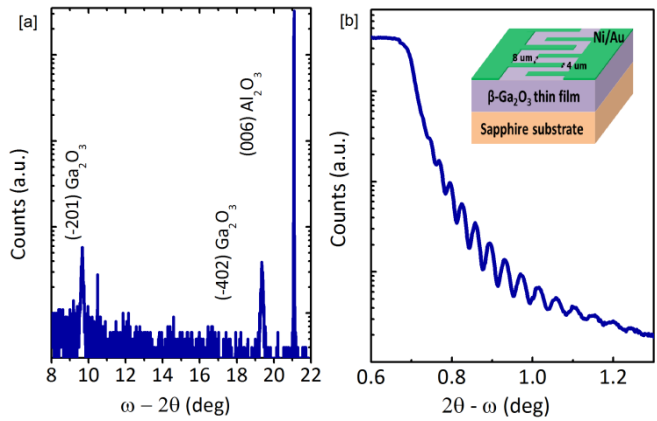

**Figure 1: (a)** X-ray diffraction and **(b)** X-ray reflectivity data of MBE-grown $Ga_2O_3$ on sapphire substrate. The inset to figure 1(b) shows schematic of the fabricated lateral MSM photodetector with interdigitated geometry.

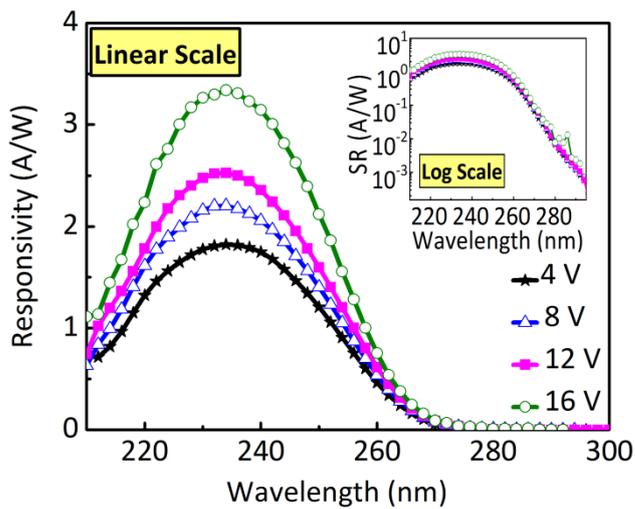

**Figure 2:** Spectral response of β-$Ga_2O_3$ MSM photodetector in linear scale at different bias voltages. The inset shows spectral response versus wavelength at different biasing voltages (Log scale).

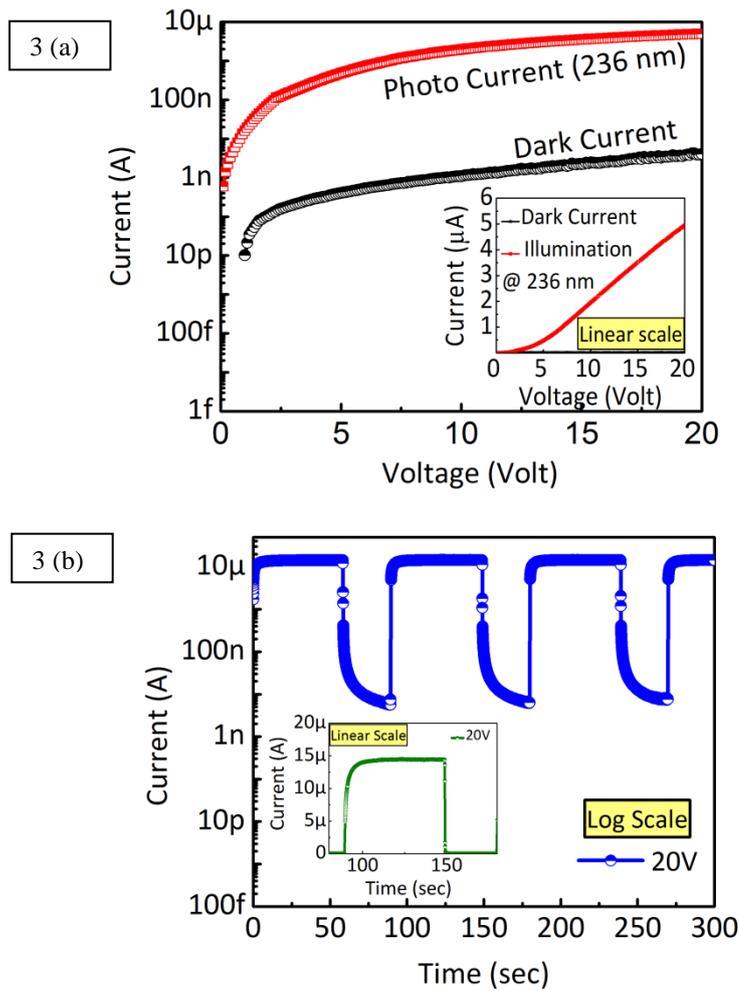

**Figure 3:** Before passivation **(a)** Photo and dark I-V characteristics at room temperature (log scale) **(b)** Time-dependent photo-response under the illumination of 236 nm at 20 V (log scale). The inset to figure 3 (a) shows photo and dark I-V characteristics in linear scale. The inset to figure 3 (b) shows rescaled transient at 20 V (linear scale).

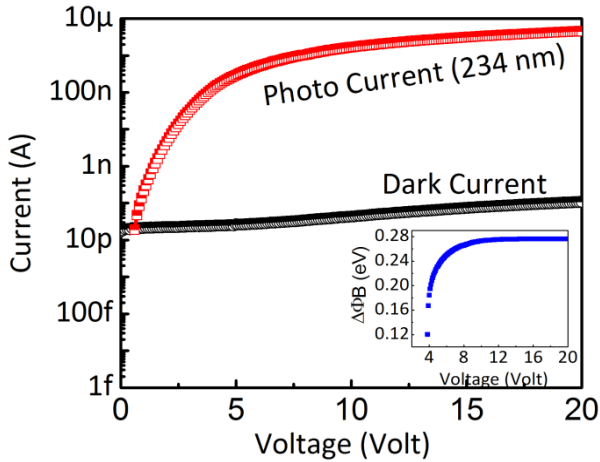

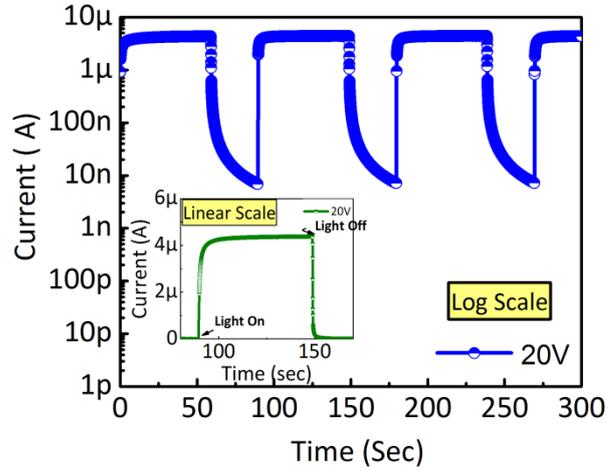

**Figure 4:** After passivation **(a)** Photo and dark I-V characteristics at room temperature (log scale) **(b)** Time-dependent photo-response under the illumination of 236 nm at 20 V (log scale). The inset to figure 4 (a) shows Schottky barrier lowering variation with the bias voltage (extracted using equation 1). The inset to figure 4 (b) shows rescaled transient at 20 V (linear scale).

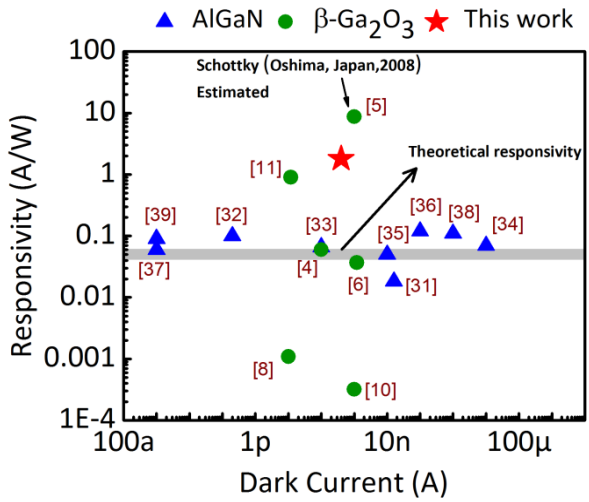

**Figure 5:** State-of-the-art plot of the responsivity versus dark current for solar-blind (230-290 nm) UV detectors based on AlGaN and β-$Ga_2O_3$. Grey line shows theoretical responsivity (Assuming external quantum efficiency ~100 %) of the deep UV photodetector (230 nm – 260 nm).